\documentclass[structabstract]{aa}  
\usepackage{amssymb}
\usepackage{graphicx}
\usepackage{epsfig}
\usepackage{subfigure}
\usepackage[authoryear]{natbib}
\usepackage[figuresright]{rotating}

\usepackage{txfonts}
\newcommand{\teff}{\ifmmode T_{\rm{eff}} \else $T_{\rm{eff}}$\fi}
\newcommand{\logg}{\ifmmode \log g \else $\log g$\fi}
\newcommand{\lL}{\ifmmode \log \frac{L}{L_{\odot}} \else $\log \frac{L}{L_{\odot}}$\fi}
\newcommand{\mdot}{$\dot{M}$}
\newcommand{\myr}{$M_{\odot}$ yr$^{-1}$}
\newcommand{\vsini}{$V$ sin$i$}
\newcommand{\vinf}{$v_{\infty}$}

\newcommand{\vmac}{$v_{\rm mac}$}

\newcommand{\kms}{km~s$^{-1}$}
\newcommand{\msun}{\ifmmode M_{\odot} \else $M_{\odot}$\fi}
\newcommand{\zsun}{\ifmmode Z_{\odot} \else $Z_{\odot}$\fi}
\newcommand{\lsun}{\ifmmode L_{\odot} \else $L_{\odot}$\fi}
\newcommand{\rsun}{\ifmmode R_{\odot} \else $R_{\odot}$\fi}
\newcommand{\qh}{\ifmmode Q_{\rm H} \else $Q_{\rm H}$\fi}
\newcommand{\qhei}{\ifmmode Q_{\ion{He}{i}} \else $Q_{\ion{He}{i}}$\fi}

\begin{document}
   \title{The two components of the evolved massive binary LZ Cep}

   \subtitle{Testing the effects of binarity on stellar evolution}

   \author{L. Mahy\inst{1}
          \and
          F. Martins\inst{2}
          \and
          C. Machado\inst{2}
          \and
          J.-F. Donati\inst{3}
          \and
          J.-C. Bouret\inst{4,5}
          }

   \offprints{L. Mahy}

   \institute{Institut d'Astrophysique et de G\'eophysique, Universit\'e de Li\`ege, B\^at B5C, All\'ee du 6 Ao\^ut 17, B-4000, Li\`ege, Belgium\\
              \email{mahy@astro.ulg.ac.be}
         \and 
             LUPM--UMR 5299, CNRS \& Universit\'e Montpellier II, Place Eug\`ene Bataillon, F-34095, Montpellier Cedex 05, France\\
         \and
             IRAP--UMR 5277, CNRS \& Univ.\ de Toulouse, 14 Av.\ E.~Belin, F--31400 Toulouse, France\\
         \and
             LAM--UMR 6110, CNRS \& Universit\'e de Provence, rue Fr\'ed\'eric Joliot-Curie, F-13388, Marseille Cedex 13, France\\
         \and
             NASA/GSFC, Code 665, Greenbelt, MD 20771, USA\\
             }

   \date{Received ...; accepted ...}

 
  \abstract
   {}
   {We present an in-depth study of the two components of the binary system LZ\,Cep in order to constrain the effects of binarity on the evolution of massive stars. }
   {We use a set of high-resolution, high signal--to-noise ratio optical spectra obtained over the orbital period of the system to perform a spectroscopic disentangling and derive an orbital solution. We subsequently determine the stellar properties of each component through an analysis with the CMFGEN atmosphere code. Finally, with the derived stellar parameters, we model the Hipparcos photometric light curve using the program NIGHTFALL to obtain the inclination and the real stellar masses.}
   {LZ\,Cep is a O\,9III+ON\,9.7V binary. It is as a semi-detached system in which either the primary or the secondary star almost fills up its Roche lobe. The dynamical masses are about 16.0 \msun\ (primary) and 6.5 \msun\ (secondary). The latter is lower than the typical mass of late--type O stars. The secondary component is chemically more evolved than the primary (which barely shows any sign of CNO processing), with strong helium and nitrogen enhancements as well as carbon and oxygen depletions. These properties (surface abundances and mass) are typical of Wolf-Rayet stars, although the spectral type is ON\,9.7V. The luminosity of the secondary is consistent with that of core He--burning objects. The preferred, tentative evolutionary scenario to explain the observed properties involves mass transfer from the secondary -- which was initially more massive-- towards the primary. The secondary is now almost a core He--burning object, probably with only a thin envelope of H--rich and CNO processed material. A very inefficient mass transfer is necessary to explain the chemical appearance of the primary. Alternative scenarios are discussed but they suffer from more uncertainties.}
   {}


   \maketitle
   \titlerunning{The two components of the evolved massive binary LZ\,Cep}
   \authorrunning{L. Mahy et al.}


\section{Introduction}
\label{s_intro}

The binary fraction of massive stars is not well constrained, but studies of young open clusters indicate that it can easily reach 60\% \citep[e.g.][]{san08}. The effects of binarity on stellar evolution depend not only on the properties of the components, but also on the mass ratio, the separation, the eccentricity. Consequently, their study is complex, requiring the exploration of a large parameter space \citep{dem11}. In the case of young systems with large separations, the components of the binary evolve independently, following single-star evolution \citep{pav05}. In evolved binary systems, the evolutionary scheme is completely different. If Roche lobe overflow happens, the surface chemistry and rotational rate of the components are deeply affected. In particular, the mass gainer is expected to show large nitrogen surface abundance enhancements \citep{langer08}. In order to better understand the physics of binary systems and to test the predictions of theoretical models, it is thus important to quantitatively analyze the properties of massive binary systems with well-constrained orbital parameters.

\begin{figure*}[ht]
\centering
\includegraphics[width=14cm, height=10cm, bb = 65 195 577 555, clip]{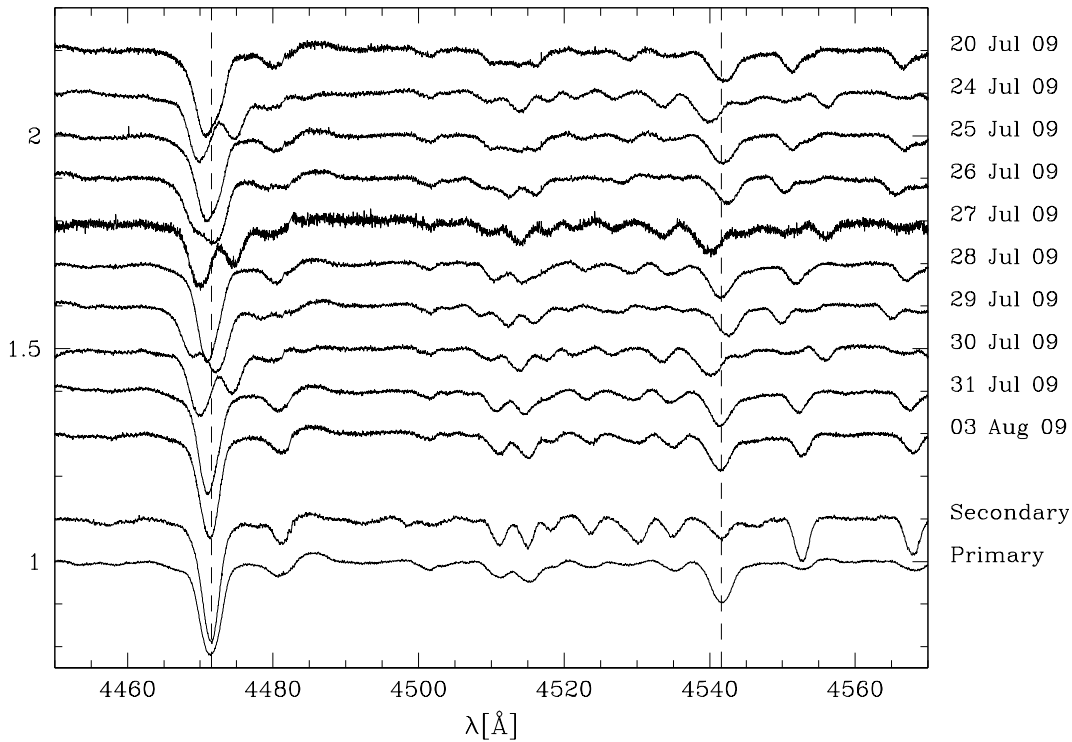}
\caption{NARVAL spectra between 4450 and 4570~\AA. The upper part of the figure shows the daily spectra. The lower two spectra are the results of the disentangling process. The two vertical dash lines correspond to the rest wavelengths of the \ion{He}{i} 4471 and \ion{He}{ii} 4542 lines}. \label{spectre}
\end{figure*}

LZ\,Cep (HD\,209481) is a double-lined spectroscopic binary system composed of a primary component with a spectral type O\,8.5III and an O\,9.5V secondary \citep{con71}. The projected rotational velocities were reported to be close to 155~\kms~and 105~\kms~by \citet{howarth91}. The period of the system was found to be 3.070508~d by \citet{rao72}, and was confirmed by \citet{howarth91} and \citet{harries98}. The eccentricity estimated by \citet{howarth91} is about 0.031. Although small, this is significant enough to not be neglected. Indeed, according to \citet{luc71}, the null hypothesis of a circular orbit can be rejected with a significance level of about 0.05. However \citet{harries98} determined an orbital eccentricity close to zero in their orbital solution ($e = 0.017\pm0.084$). It thus appears that the circularization of the system is not clearly established. From the photometric point of view, LZ\,Cep is not an eclipsing system but ellipsoidal variations are clearly visible in the light curve. The photometric analysis indicates that this binary is an evolved system \citep{hill76} with a secondary component that fills up its Roche lobe. Hence, the system is most likely in a configuration of post case~A mass transfer i.e., a system where the mass transfer happened while both components are still in a hydrogen-burning core phase \citep{howarth91}.

LZ\,Cep is thus a target of choice for the study of the effects of binary interactions and their impacts on massive star evolution. The present paper is organized as follows. Sect.~\ref{s_obs} presents the journal of observations and the data-reduction technique. Then, we devote the Sect.~\ref{s_orb} to the determination of the orbital solution and the spectroscopic classification. Using the radial velocities obtained from the orbital solution, we apply a disentangling programme to separate the individual spectrum of both components. In Sect.~\ref{s_spec}, we apply the CMFGEN atmosphere code on the separated spectra to determine the main stellar and wind parameters as well as the surface abundances of each star. These parameters provide us with information about the stars' evolutionary status. The last step in our investigation is reported in Sect.~\ref{s_light} and consists in the analysis of the Hipparcos light curve. It provides the system inclination and consequently the real masses of the stars. All the parameters and results determined in the present paper are discussed in Sect.~\ref{s_disc}. Finally, we present the conclusions in Sect.~\ref{s_conc}.


\section{Observations and data reduction}
\label{s_obs}

LZ\,Cep was observed with the spectropolarimeter NARVAL mounted on the T\'elescope Bernard Lyot (TBL) located at the Pic du Midi observatory in the french Pyrenees. Table~\ref{tab_obs} gives the journal of observations. A total of ten spectra over a period of 14~d were obtained between July 20$^{\rm{th}}$ and August 3$^{\rm{rd}}$ 2009 (Fig~\ref{spectre}). This corresponds to more than 3 orbital periods of the system. The signal--to--noise ratios per 2.6 \kms\ velocity bin are between 300 and 1300 depending on weather conditions and wavelength. A spectral resolution of 65000 is achieved. The data were automatically reduced with the Libre ESpRIT package \citep{donati97} installed at TBL. An exhaustive description of the reduction procedure is given in \citet{donati97} and we refer to this paper for information.

\begin{table*}
\begin{center}
\caption{Journal of the observations. } \label{tab_obs}
\begin{tabular}{llccrr}
\hline
Date  & HJD & exposure time & SNR & $RV_{\rm{P}}$ & $RV_{\rm{S}}$\\
      & [d] & [s]           &     & [\kms] & [\kms]\\
\hline
20 Jul 2009      & 2455033.48549 & 2700              &   330--750  &   26.7  &   $-$100.1\\
24 Jul 2009      & 2455037.48790 & 3300              &   630--1320 &  $-$96.6  &    211.8\\
25 Jul 2009      & 2455038.49492 & 2700              &   480--1000 &   26.3  &    $-$95.8\\
26 Jul 2009      & 2455039.49494 & 2700              &   430--900  &   45.7  &   $-$153.9\\
27 Jul 2009      & 2455040.49228 & 2700              &   160--350  &  $-$98.4  &    206.6\\
28 Jul 2009      & 2455041.48290 & 2700              &   550--1180 &   10.0  &    $-$66.0\\
29 Jul 2009      & 2455042.49901 & 2700              &   640--1330 &   52.1  &   $-$179.1\\
30 Jul 2009      & 2455043.47908 & 2700              &   560--1150 &  $-$95.8  &    199.2\\
31 Jul 2009      & 2455044.49895 & 2700              &   570--1230 &   $-$4.4  &  $-$40.3\\
03 Aug 2009      & 2455047.46335 & 2700              &   500--1080 &  $-$23.8  &      5.3\\
\end{tabular}
\tablefoot{
The HJD is measured at mid exposure. The signal-to-noise ratio (SNR) is given in the wavelength range 4000--6700 \AA\ and depends on the exact wavelength. Radial velocities are obtained by cross-correlation and have an error of 5 \kms.}
\end{center}
\end{table*}


\section{Disentangling and orbital solution}
\label{s_orb}
Previous analyses by \citet{howarth91} and \citet{harries98} focused on the orbital elements of the system. In the present paper, we go one step further and study the properties of the individual components.
The first step is to separate the spectra of each star. We applied an iterative disentangling method which consists in using alternatively the spectrum of one component (shifted by its radial velocity) and to subtract it from the observed spectra to determine the mean spectrum of the other component. The separated spectra, corrected by the brightness ratio (see below), are shown in Fig.\ \ref{spectre} (the two bottom spectra) for the range 4450--4570~\AA~and in Figures~\ref{fit_opt_1} and \ref{fit_opt_2} (black lines) for a larger spectral band. This disentangling method, inspired by \citet{gl06}, requires the knowledge of approximative radial velocities as input. We first measured these radial velocities by fitting two Gaussian profiles to the observed spectra. The cross-correlation technique was subsequently used by the disentangling program to refine the radial velocities from the disentangled spectra. This method ensures that radial velocities are obtained at all orbital phases, including those where the lines are heavily blended. The cross-correlation masks were built on a common basis of the \ion{Si}{IV}~4089--4116, \ion{He}{I}~4388, 4471, 4713, 4921, 5015, 5876, 6678, \ion{C}{IV}~5801--5812, \ion{O}{III}~5592 and H$\alpha$ lines. For the primary component, we added the \ion{He}{II}~4200, 4542, 5412 lines indicating, by the way, its earlier spectral type \citep{con71}. The list of the radial velocities obtained by cross-correlation is reported in Table~\ref{tab_obs}. 

\begin{figure}[ht]
\centering
\includegraphics[width=8cm]{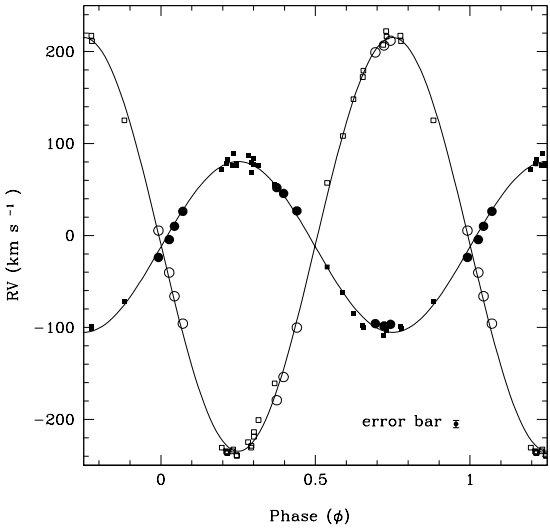}
\caption{Radial velocities as a function of orbital phase. The solid (open) symbols are for the primary (secondary). The circles are the data of the present paper whilst the squares are the data of \citet{harries98}.}\label{sol_orb}
\end{figure}

We checked these radial velocities by computing the orbital solution of LZ\,Cep. For this purpose, we took into account the well-known period of the system ($P_{\mathrm{orb}}=3.070507$~d, \citealt{pet62}) and we used the Li\`ege Orbital Solution Package (LOSP, Sana\,\&\,Gosset, A\&A, submitted). This software is based on the generalization of the SB1 method of \citet{wol67} to the SB2 case along the lines described in \citet{rau00} and \citet{san06a}. We assigned a weight of 1.0 to all the measurements. Figure~\ref{sol_orb} represents the fitted radial velocity curves as a function of the phase. We added, for comparison, the radial velocities measured from the {\it IUE} spectra by \citet{harries98}. In order to compare both datasets, it was necessary to shift the radial velocities of \citet{harries98} by $-9.4$ \kms\ to obtain the same systemic velocity than that measured for our dataset. We represent these data by squares in Fig.~\ref{sol_orb}. $T_0$ is the phase of the conjunction with the primary star in front. We tried to include a non-zero eccentricity in the orbital solution but the fit residuals are smaller for a zero eccentricity, indicating that the system is circularized. Previous studies \citep{howarth91,harries98} have shown that this system could be in a post case A mass transfer phase (see also Sect.\ \ref{s_light}). It is therefore likely that the tidal interactions have circularized that system. Other differences between our solution and that of \citet{howarth91} are notably a smaller value of $K_2$, smaller minimum masses and a higher mass ratio. The orbital parameters are listed in the upper part of Table~\ref{tab_orb}.

Finally, in the frame of the study of wind parameters (see Sect.~\ref{s_spec}), we also decided to separate the {\it IUE} spectra in the range 1200--1800~\AA. We thus determined, from the orbital solution, the values of the radial velocities corresponding to the {\it IUE} spectra. We fixed these radial velocities in the disentangling program to obtain the resulting spectra shown in Fig.~\ref{fit_UV_1} (black line). Unfortunately, the disentangled spectrum of the secondary is too noisy to be used and to constrain the wind parameters.

In order to pinpoint the nature of the components, we determined their spectral types. We measured the equivalent widths by fitting Gaussian functions to the line profiles. We did that for the most deblended observed spectra and from the disentangled spectra. The spectral-type determination is based on the quantitative classification criteria for O-type stars proposed by \citet{con71}, \citet{con73} and \citet{mat88,mat89}. We derived, from the ratios between the appropriate lines, the spectral classifications O\,9III and ON\,9.7V for the primary and the secondary star, respectively. The ``N'' classification of the secondary comes from the strength of the nitrogen lines -- e.g. \ion{N}{ii} 4037--41--43, \ion{N}{ii} 4601--07--13 and \ion{N}{iii} 4523--30--34. These results confirm the conclusion of \citet{hil74} who reported that the secondary is of different luminosity class than the primary component. Otherwise, the spectral types are slightly different from those quoted in the literature. Indeed, \citet{pet62} quoted LZ\,Cep as a binary system composed by two O\,9.5V stars whilst \citet{con71} determined a spectral type O\,8.5III for the primary and O\,9.5V for the secondary component.

In a second step, we estimated the light ratio. For that purpose, we measured, for both stars, the equivalent widths of the \ion{He}{I}~4026, 4143, 4388, 4471, 5876, \ion{He}{II}~4200, 4542, \ion{Si}{IV}~4089, \ion{O}{III}~5592 lines and we then computed the ratio between these lines and the corresponding theoretical average values reported in \citet{con71}, \citet{con73} and \citet{con77}. Due to the absence of O\,9.7 stars in the latter three papers, we used an O\,9.5 star as reference for the secondary. We find that the primary star is about 1.85 times brighter than the secondary ($l_1/l_2 = 1.85\pm0.1$). Alternatively we also measured the equivalent widths of the above-quoted lines in synthetic spectra computed with a similar effective temperature and \logg\ to those of the primary and secondary components estimated (see Sect.~\ref{s_spec}). With these synthetic spectra, we obtained a light ratio of about $l_1/l_2 = 1.91$ which is consistent with our previous value ($l_1/l_2 = 1.85\pm0.1$). Consequently, we adopt in the following a light ratio of about $l_1/l_2 = 1.85\pm0.1$. \citet{pet62} reported a value of $l_2/l_1 = 0.66\pm0.1$ which corresponds to $l_1/l_2 = 1.52\pm0.3$ (the method they used is not described though). Another estimate was made by \citet{harries98}. Although their method appears relevant, the result seems questionable, probably due to the confusion between the primary and secondary stars. Indeed, the authors reported a value of about $l_1/l_2 = 0.65\pm0.1$. However, when we compare the equivalent widths computed by \citet{harries98} with our measurements, no significant difference appears, indicating that the equivalent widths of the primary are larger than those measured on the secondary. We thus assume that the real determination of \citet{harries98} is $l_2/l_1 = 0.65\pm0.1$ which corresponds to $l_1/l_2 = 1.54\pm0.3$. Consequently, the discrepancy between our measurement and that of \citet{harries98} and \citet{pet62} probably comes from the sample of spectral lines that is larger in our analysis, rendering more accurate the determination of the light ratio \footnote{If we rely on the same spectral lines as \citet{harries98} for the determination, we obtain a brightness ratio only slightly different from their estimate.}. As a comparison, if we use a smaller brightness ratio ($l_1/l_2 \sim 1.54$), we obtain absolute luminosities of $\log(L_1/L_{\odot}) = 5.08$ and $\log(L_2/L_{\odot}) = 4.74$. The effect on the determination of the abundances is however small and does not exceed the error bars quoted in Sect.~\ref{s_spec}.

Adopting our brightness ratio, we computed the luminosity of each component. The magnitude and color of the system are $V=5.55$ and $(B-V)=0.06$ \citep{mai04}. The visual extinction directly follows from $(B-V)_0 = -0.27$ \citep[appropriate for late O stars,][]{mar06}. With these photometric parameters and assuming a distance modulus for LZ\,Cep equal to $DM = 10.06^{+0.46}_{-0.38}$ \citep[][and references therein]{sota2011}, consistent with the value of 10.12 given by \citet{pet62}, we found a visual absolute magnitude of the entire binary system to be equal to $M_V = -5.53^{+0.38}_{-0.46}$. We derived, by using the brightness ratio, $M_{V_{\mathrm{P}}} = -5.06^{+0.38}_{-0.46}$ and $M_{V_{\mathrm{S}}} = -4.39^{+0.38}_{-0.46}$ and with the bolometric corrections corresponding to the spectral types of both components \citep{mar06}, we obtained absolute luminosities of $\log(L_1/L_{\odot}) = 5.11^{+0.19}_{-0.16}$ and $\log(L_2/L_{\odot}) = 4.69^{+0.19}_{-0.16}$ (more details on the method are given in \citealt{mah10}). As a comparison, if we took into account a smaller brightness ratio ($l_1/l_2 \sim 1.54$), we would obtain absolute luminosities of $\log(L_1/L_{\odot}) = 5.08$ and $\log(L_2/L_{\odot}) = 4.74$. Accordingly, the brightness ratio computed in the present analysis does not affect significantly the determination of the results since these values remain within the measured error bars.

\begin{table}
\begin{center}
\caption{Orbital solution and stellar parameters. } \label{tab_orb}
\begin{tabular}{lrr}
\hline
                    & Primary        & Secondary \\
\hline
$P$ [d] & \multicolumn{2}{c}{3.070507 (fixed)}\\
$e$ &  \multicolumn{2}{c}{0.0 (fixed)}\\
$T_0$ [HJD -- 2450000]  & \multicolumn{2}{c}{5032.019$\pm$0.002}\\
$q (M_1/M_2)$ & \multicolumn{2}{c}{2.53$\pm$0.05}\\
$\gamma$ [\kms] & $-$11.81$\pm$0.91 & $-$11.40$\pm$1.20\\
$K$ [\kms] & 88.72$\pm$1.02 & 224.48$\pm$2.58\\
$a \sin i$ [$R_{\odot}$] &5.38$\pm$0.06 & 13.61$\pm$0.16\\
$M \sin^3 i$ [\msun] & 7.00$\pm$0.18 & 2.77$\pm$0.05\\
rms [\kms] & \multicolumn{2}{c}{2.9198}\\
\hline
\teff\ [K]          & 32000$\pm$1000      & 28000$\pm$1000 \\
\lL\                & 5.11$^{+0.19}_{-0.16}$ & 4.69$^{+0.19}_{-0.16}$ \\
\logg\              & 3.5$\pm$0.1         & 3.1$\pm$0.1 \\
$M_{spec}$ [\msun]   & 15.9$^{+9.8}_{-8.4}$   & 4.1$^{+2.4}_{-2.1}$ \\
$M_{ev}$ [\msun]    & 25.3$^{+6.2}_{-4.0}$    & 18.0$^{+2.4}_{-2.5}$ \\
radius [$R_{\odot}$] & 11.7$^{+3.3}_{-2.7}$    &  9.4$^{+2.6}_{-2.2}$\\
He/H                & 0.1$\pm$0.02         &  0.4$\pm$0.1\\
C/H [$\times\ 10^{-4}$]& 1.0$\pm$0.5          &  0.3$\pm$0.2\\
N/H [$\times\ 10^{-4}$]& 0.85$\pm$0.2         &  12.0$\pm$2.0\\
O/H [$\times\ 10^{-4}$]& 3.0$\pm$0.5          &  0.5$\pm$0.3\\
\vsini\ [\kms]      & 130$\pm$10                  & 80$\pm$10 \\
\vmac\ [\kms]       & 40$\pm$5                   & 44$\pm$5 \\
\mdot\ [10$^{-8}$\myr] & 1.0$\pm0.3$        & -- \\
\vinf\ [\kms]       & 1800$\pm$100         & -- \\
\end{tabular}
\tablefoot{ The given errors correspond to 1--$\sigma$. The solar abundances for the chemical elements quoted here are He/H $= 0.1$, C/H $ = 2.45\times10^{-4}$, N/H $ = 0.60\times10^{-4}$, O/H $ = 4.57\times10^{-4}$, respectively.}
\end{center}
\end{table}


\section{Spectroscopic analysis}
\label{s_spec}

\begin{figure}[]
\centering
\includegraphics[width=8cm]{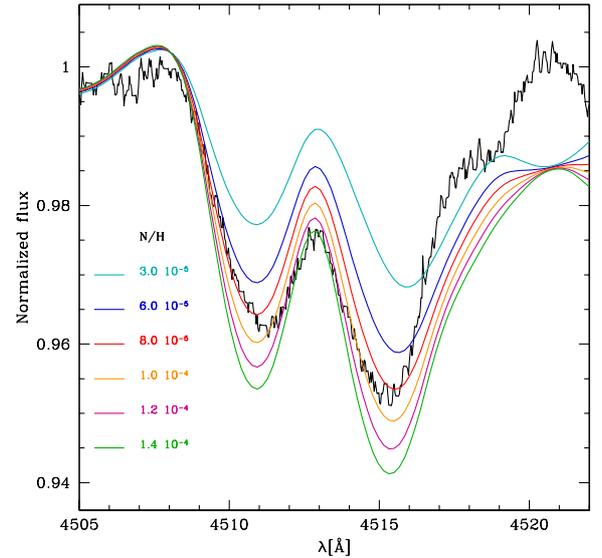}
\caption{Illustration of the N/H determination for the primary component of LZ\,Cep. Colored lines are models and the black line is the disentangled spectrum. We estimate that the best fit abundance has N/H slightly above 8.0$\times 10^{-5}$ and that models with $\rm{N/H} =6.0 \times 10^{-5}$ and $\rm{N/H}=1.0\times 10^{-4}$ correspond to the extrema of the allowed N/H values.}\label{fit_N_prim}
\end{figure}

The disentangled optical spectra of both components have been analyzed separately with the atmosphere code CMFGEN \citep{hm98}. The UV spectrum of the primary was also used. CMFGEN produces non-LTE models including winds and line-blanketing. The rate equation are solved iteratively together with the radiative transfer equation to provide the occupation numbers of energy levels and the radiation field. The computations are done in spherical geometry and in the co-moving frame. The temperature structure results from the radiative equilibrium condition. The velocity structure is an input of the calculation. In our case, we used TLUSTY hydrostatic structures in the photosphere connected to a $\beta$--velocity law for the wind part. The OSTAR2002 grid of TLUSTY models was used \citep{lh03}. The density structure directly follows from the mass-conservation equation. Once the atmosphere model is obtained, a formal solution of the radiative transfer equation is performed to yield the synthetic spectrum. For that, accurate line profile including Stark broadening and a depth variable microturbulence are used. The latter varies from 10 \kms\ in the photosphere to 0.1$\times$\vinf\ in the outer atmosphere (\vinf\ being the terminal velocity). The spectrum is subsequently compared to the observational data to constrain the main parameters. In practice, we proceed as follows:

\begin{itemize}

\item {\it effective temperature}: \teff\ is determined from the ionization balance of He. The relative strength of \ion{He}{i} and \ion{He}{ii} directly depends on the temperature. We used \ion{He}{i} 4471 and \ion{He}{ii} 4542 as the main indicators. Other lines were also checked: \ion{He}{i} 4026, \ion{He}{i} 4389, \ion{He}{i} 4713, \ion{He}{i} 5876, \ion{He}{ii} 4200, \ion{He}{ii} 4686, \ion{He}{ii} 5412. An accuracy of $\pm$1000 K in the \teff\ determination was achieved.

\item {\it gravity}: larger \logg\ broadens the width of the Balmer lines wings. We thus used H$\beta$, H$\gamma$ and H$\delta$ as the main gravity diagnostics. The final \logg\ are accurate to 0.1 dex.

\item {\it luminosity}: the knowledge of the effective temperature, of the absolute V magnitude and of the extinction provides a direct estimate of the luminosity as described in Sect.\ \ref{s_orb}. The uncertainty on the luminosity determination depends mainly on the uncertainty on the distance (see Sect.\ \ref{s_orb}).

\item {\it surface abundances}: The surface abundances were determined from the strength of He, C, N and O lines. He lines of all ionization states are stronger when the helium content is higher. This effect is different from a change in \teff\ which changes the relative strengths of lines from successive ionization states. We have thus used the same lines as for the effective temperature determination to constrain the He content. For nitrogen, we mainly relied on several \ion{N}{ii} lines between 4000 and 4700 \AA~as well as the \ion{N}{iii} lines in the 4500--4520 \AA\ range.  The main carbon indicators were the \ion{C}{iii} lines around 4070 \AA\ and \ion{C}{iii} 5696 \AA. For oxygen, we relied on the numerous \ion{O}{ii} lines between 4000 and 4600 \AA\ as well as on \ion{O}{iii} 5592 \AA. The typical uncertainty on the CNO surface abundance is 50\%. To estimate these uncertainties, we ran models with element contents bracketing the preferred value and compared the corresponding spectra to the observed line profiles. When a clear discrepancy was seen on all lines of the same element, we adopted the corresponding abundance as the maximum/minimum value of the possible chemical composition. An illustration is shown in Fig. \ref{fit_N_prim}. As a test, we conducted the same analysis for the primary with four additional sets of models with the following sets of parameters: $\teff=31000$~K and $\logg=3.5$; $\teff=33000$~K and $\logg = 3.5$; $\teff=32000$~K and $\logg = 3.4$; $\teff = 32000$~K and $\logg = 3.6$. We then estimated the best N/H abundance for each set of models (including also the best fit combination, $\teff=32000$~K and $\logg = 3.5$) by means of a chi--square method. We then computed the average and 1--sigma dispersion of this set of five N/H values and found $\rm{N/H} = 8.5\pm0.18 \times 10^{-5}$. This is perfectly consistent with the value we estimate from Fig.\ \ref{fit_N_prim}. For the other elements and the secondary, we thus proceeded as described above, running models with various abundances for the best fit parameters (\teff,~\logg).

\item {\it mass-loss rate}: we used the UV resonance lines of the primary to estimate \mdot. \ion{C}{iv} 1538--1545 and \ion{N}{v} 1240 were the main diagnostics. The uncertainty on \mdot\ is 0.1 dex. For the secondary, the UV spectrum was not good enough to allow such a determination but in order to compute the atmosphere model associated to this star, we took a \mdot~$ = 3\times 10^{-8}$~\myr. We also used homogeneous models (i.e. no clumping was included), which provided a good fit of the UV lines. As a consequence, our determination should be seen as an upper limit. We also included X--ray emission so as to have a canonical value of $\log \frac{L_{X}}{L_{bol}}=-7$ \citep{san06b}. This turned out to be crucial to reproduce the \ion{N}{v} 1240 profile. 

\item {\it terminal velocity}: the blueward extension of the P--Cygni profiles of the primary UV spectrum are directly related to \vinf. We thus used them to estimate this parameter. \vinf\ is derived to within 100 \kms. 

\item {\it rotational and macroturbulent velocity}: we used the Fourier transform formalism \citep{sim07} to derive the projected rotational velocity (\vsini). The macroturbulent velocity was obtained by convolving the theoretical line profiles with a Gaussian profile to best reproduce the shape of \ion{He}{i} 4713. 

\end{itemize}  

\begin{figure*}[ht]
\centering
\includegraphics[width=18cm]{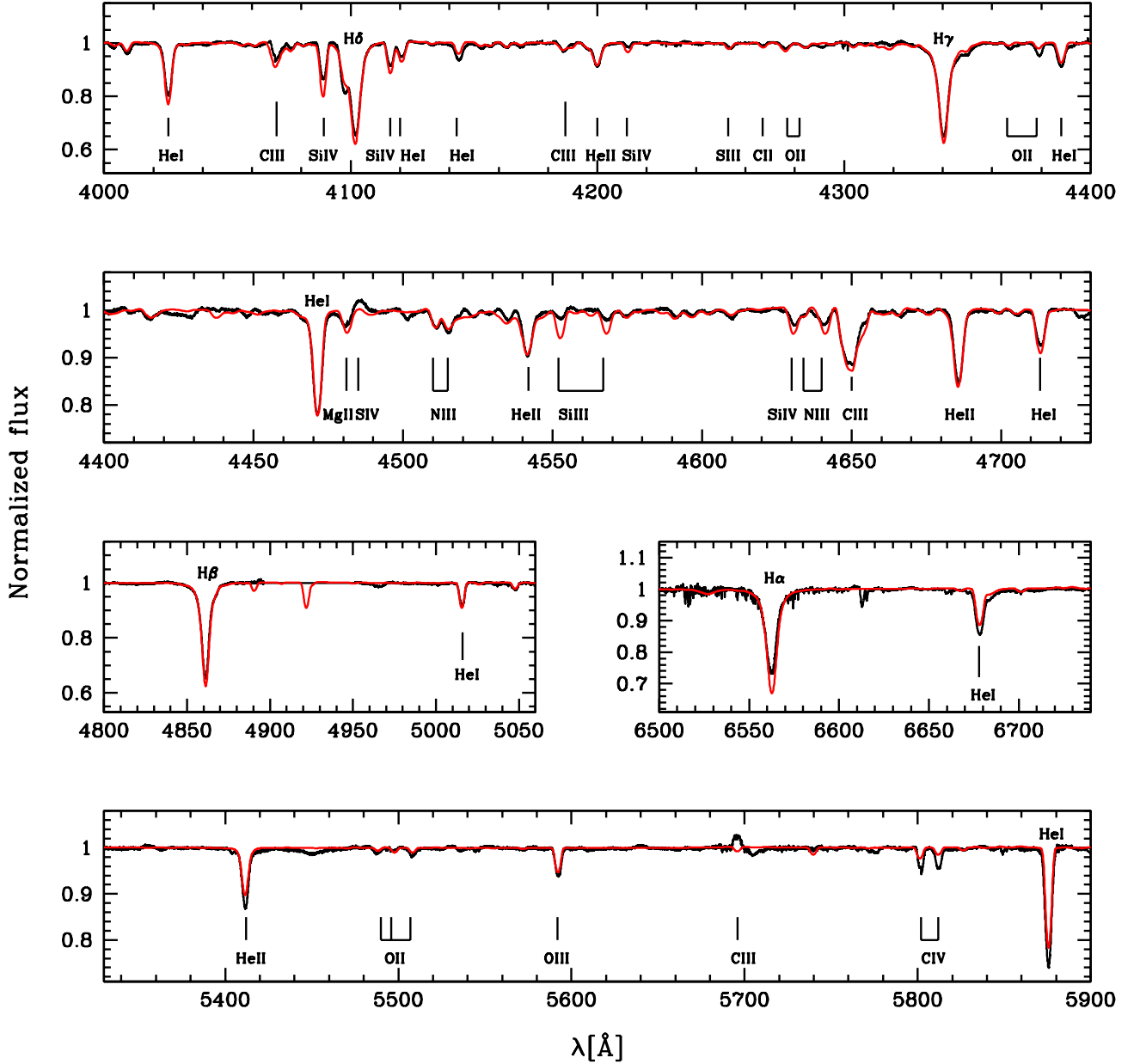}
\caption{Best CMFGEN fit (red) of the disentangled optical spectrum (black) of the primary component of LZ\,Cep. No data are available in the immediate vicinity of \ion{He}{i} 4921. }\label{fit_opt_1}
\end{figure*}

\begin{figure*}[ht]
\centering
\includegraphics[width=18cm]{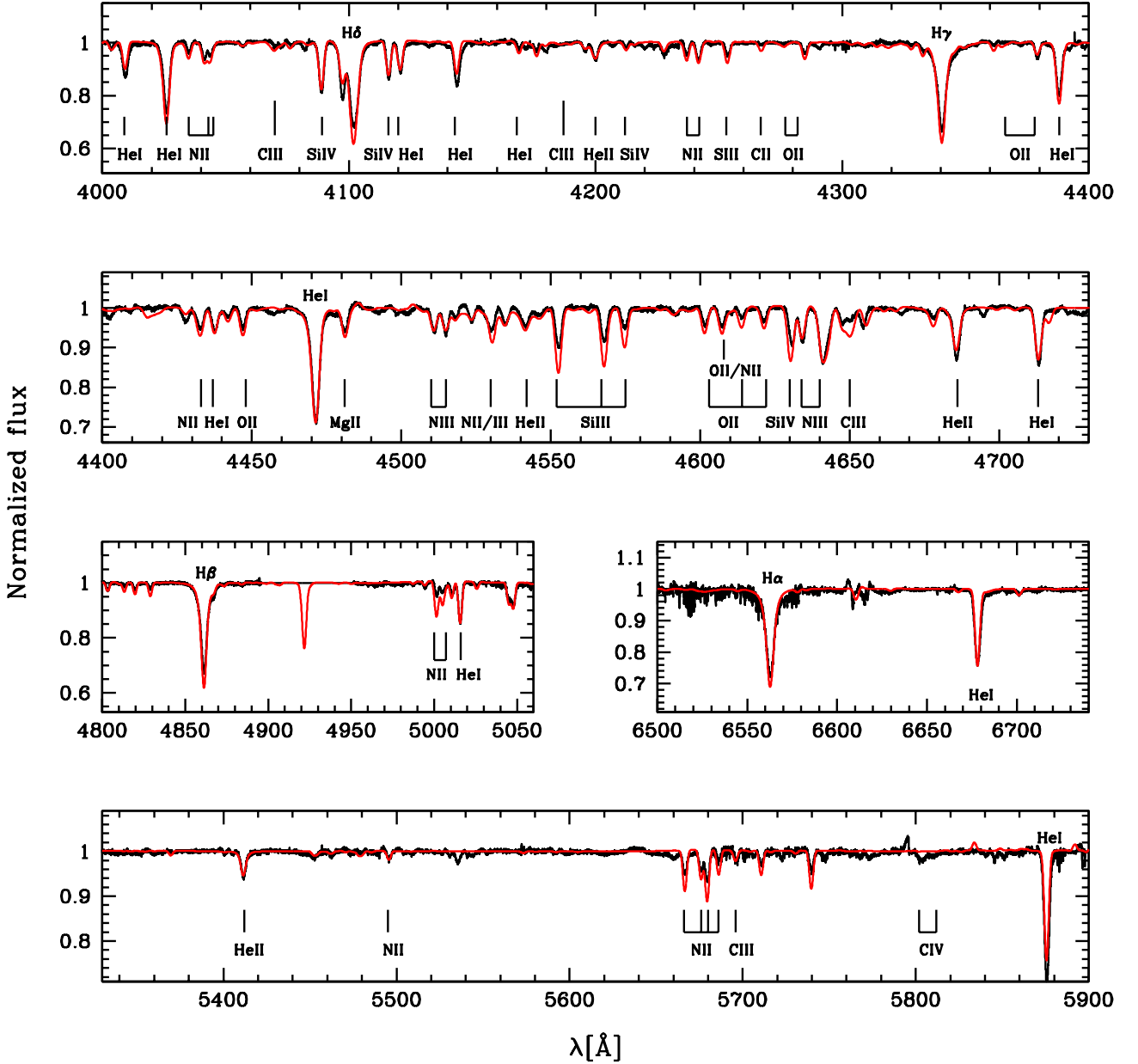}
\caption{Same as for Fig.~\ref{fit_opt_1} but for the secondary component of LZ\,Cep.  No data are available in the immediate vicinity of \ion{He}{i} 4921.}\label{fit_opt_2}
\end{figure*}

\begin{figure}[]
\centering
\includegraphics[width=8cm]{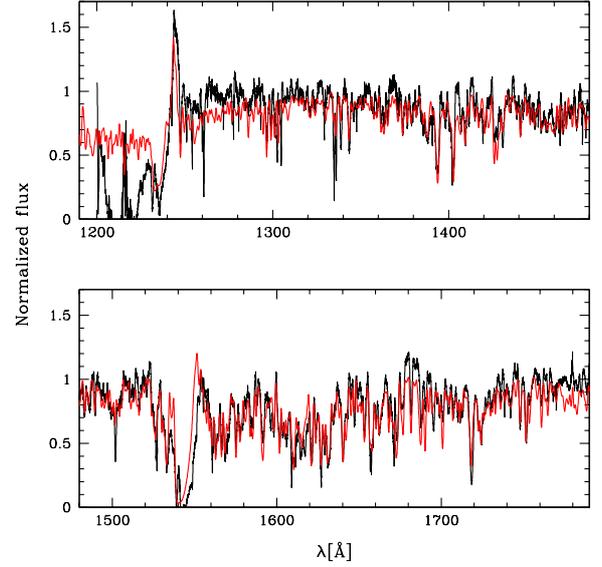}
\caption{Best CMFGEN fit (red) of the disentangled UV spectrum (black) of the primary component of LZ\,Cep.  }\label{fit_UV_1}
\end{figure}

The best fits of both components are shown in Fig.\ \ref{fit_opt_1} and Fig.\ \ref{fit_opt_2}. The fit of the UV spectrum of the primary is shown in Fig.\ \ref{fit_UV_1}. The derived parameters are gathered in Table \ref{tab_orb} (bottom part). The quality of the fit is on average very good. Only a few features are not reproduced. This is the case of the \ion{Si}{iii} lines around 4550~\AA. Increasing \teff\ leads to a better fit, but in that case the other diagnostics are less well reproduced. This problem is frequently observed in our models, and the origin is not known at present. Hence, we decided not to rely on these lines in the present study.

\section{Light curve}
\label{s_light}

To better understand the binary system LZ\,Cep, and in particular, its geometry, we also performed an analysis of the Hipparcos light curve.
This investigation was made with the NIGHTFALL programme\footnote{for more details, see the Nightfall User Manual by \citet{wic98} available at the URL: \texttt{www.hs.uni-hamburg.de/DE/Ins/Per/Wichmann/Nightfall.html}}. This software is based on a generalized Wilson-Devinney method assuming a standard Roche geometry. We performed a fit of the light curve by minimizing the free parameters. We fixed the effective temperatures to 32000~K and 28000~K, for the primary and the secondary star, respectively, as determined by the CMFGEN analysis (see Sect.\ \ref{s_spec}). The mass ratio and the orbital period are those obtained from the orbital solution (see Table\ \ref{tab_orb}). We stress that the zero phase of the light curve fit corresponds to the zero phase of the orbital solution. 

\begin{figure}[ht]
\centering
\includegraphics[width=8cm, bb = 36 392 551 692, clip]{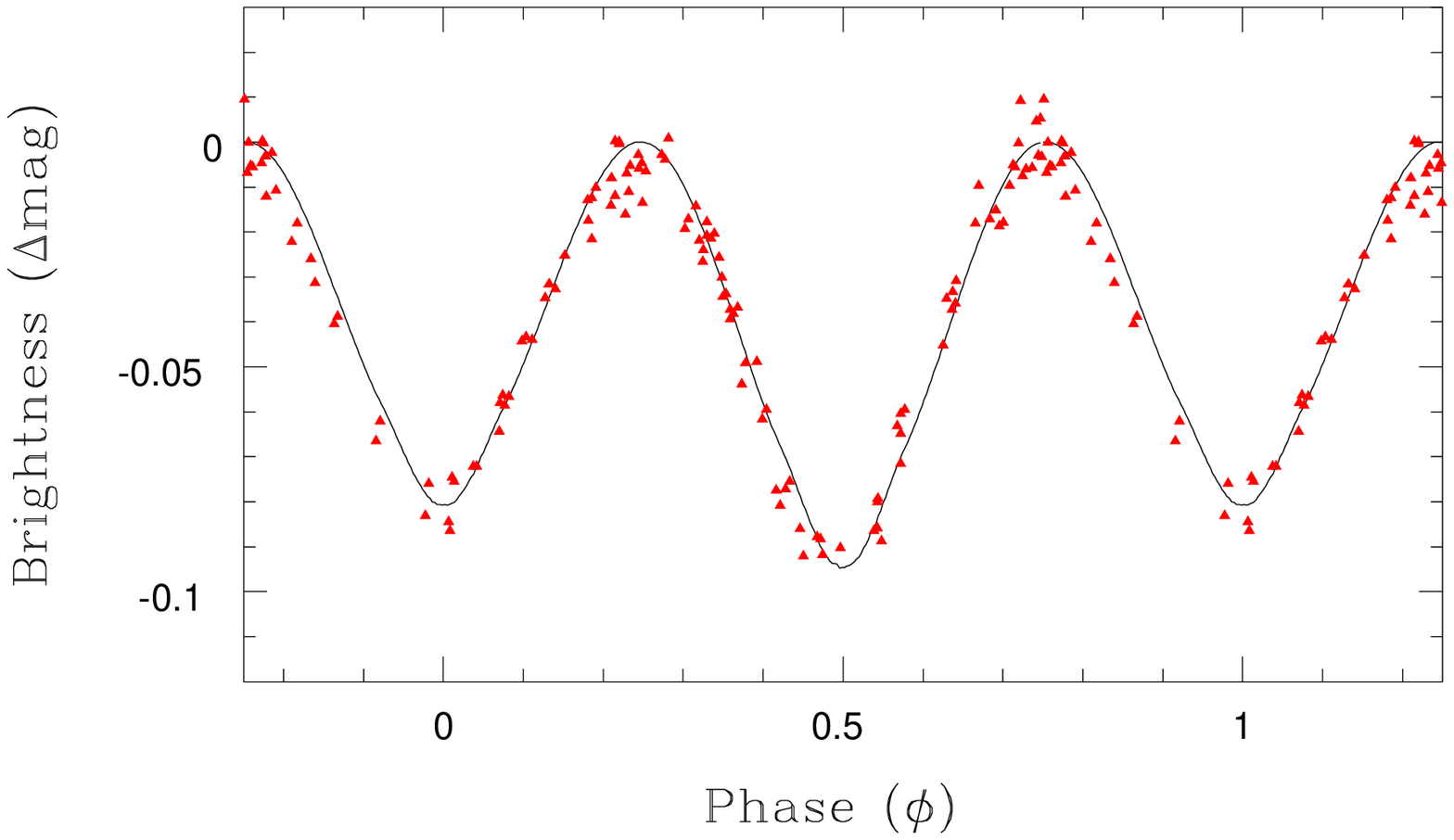}
\includegraphics[width=8cm, bb = 0 178 454 430, clip]{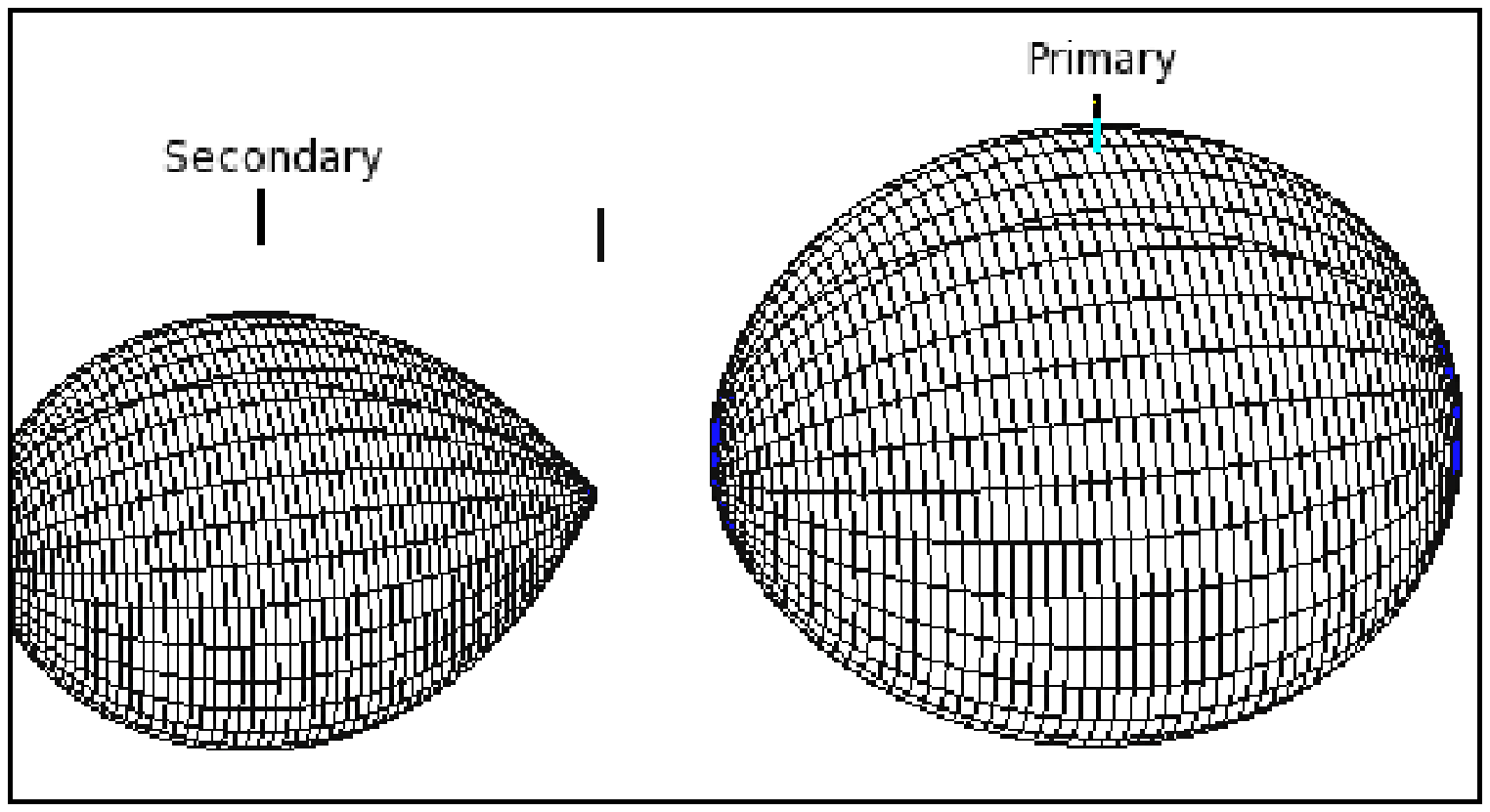}
\caption{{\it Top:} Photometry of LZ\,Cep from Hipparcos satellite. The red triangles are the observational data. A period of $\sim$3.07~d is used. The solid line corresponds to the theoretical light curve fit with the parameters listed in Table\ \ref{tab_lc}. {\it Bottom:} A 3D view of both stars when the secondary star fills up its Roche lobe.}\label{fit_lc}
\end{figure}

The light curve (Fig.~\ref{fit_lc}) does not exhibit any eclipses but ellipsoidal variations are clearly visible. The depth of the secondary minimum is very similar to that of the primary minimum. \citet{howarth91} considered two different models to explain the geometry of this binary system: a first one where both stars fill their Roche lobe, and a second one with a semi-detached configuration. The emphasis of their analysis was put on the second model with the secondary star filling its Roche lobe.

\begin{table}
\begin{center}
\caption{Parameters fitted from the Hipparcos light curve.} \label{tab_lc}
\begin{tabular}{lrr}
\hline
Parameters & Sol~1 & Sol~2\\
\hline
$i~[^{\circ}]$ & $50.1^{+2.1}_{-1.5}$ & $48.1^{+2.0}_{-0.7}$\\[1pt]
$q~[M_1/M_2] $ & $2.53$ (fixed) & $2.53$ (fixed)\\
Filling factor primary [\%] & $98.0^{+1.2}_{-1.0}$ & $94.8^{+2.0}_{-3.0}$\\[1pt]
Filling factor secondary [\%] & $87.0^{+3.0}_{-5.0}$ & $98.6^{+0.3}_{-1.5}$\\[1pt]
\teff$_{\textrm{,p}}$ [K]  & $32000$ (fixed) & $32000$ (fixed) \\
\teff$_{\textrm{,s}}$ [K]  & $28000$ (fixed) & $28000$ (fixed)\\
$M_{\textrm{p}}$ [\msun] & $15.5^{+1.0}_{-1.0}$ & $16.9^{+1.0}_{-1.0}$ \\[1pt]
$M_{\textrm{s}}$ [\msun] & $6.1^{+1.0}_{-1.0}$ & $6.7^{+1.0}_{-1.0}$ \\[1pt]
$R_{pole\textrm{,p}}$ [\rsun]& $10.5^{+1.2}_{-1.2}$ & $10.5^{+1.2}_{-1.2}$\\[1pt]
$R_{pole\textrm{,s}}$ [\rsun]& $6.1^{+1.2}_{-1.2}$ & $7.1^{+1.2}_{-1.2}$ \\[1pt]
$R_{equ\textrm{,p}}$ [\rsun]& $13.1^{+1.2}_{-1.2}$ & $12.4^{+1.2}_{-1.2}$\\[1pt]
$R_{equ\textrm{,s}}$ [\rsun]& $6.9^{+1.2}_{-1.2}$ & $9.3^{+1.2}_{-1.2}$ \\[1pt]
\end{tabular}
\tablefoot{The index 'p' ('s') refers to the primary (secondary). $R_{pole}$ is the polar radius, and $R_{equ}$ the equatorial radius.}
\end{center}
\end{table}

 In our analysis, the NIGHTFALL programme also gives us two situations but, at each time, converges towards a semi-detached system. The first possibility is a situation in which the primary component fills up its Roche lobe (Sol~1 in Table~\ref{tab_lc}). This configuration favors an inclination of about $50.1^{\circ}$. The primary star would fill up its Roche lobe at about 98\% whilst the secondary component at about 87\%. The second possible configuration is a system where the secondary star fills up its Roche lobe (Sol~2 in Table~\ref{tab_lc}). This configuration is closer to that found in the literature. Under this assumption, we found an inclination of about $48.1^{\circ}$. The volumes of the Roche lobes are filling up at about 95\% and 99\% for the primary and the secondary components, respectively. As \citet{howarth91} already mentioned and no matter the chosen configuration, the inclination is well between 45$^{\circ}$ and 55$^{\circ}$ but, with the present investigation, we note that we improved, with more accuracy, the true value of the inclination. In both configurations, we included the reflection effects which are not negligible when one star fills out a large fraction of its Roche lobe. We also note that there is no photometric evidence of a non-zero eccentricity. 

The error bars given in Table~\ref{tab_lc} are established by exploring the parameter space. We fixed one parameter and let the others vary to reach the minimum of the $\chi^2$. The error bars were then determined for a variation of the $\chi^2$ corresponding to a 90\% confidence level and the appropriate number of freedom degrees (for more details, see \citealt{lin09}).

Both models are really close to each other in terms of parameters values and it is difficult to favor one configuration without any eclipse and with only the light curve fit. We can only safely conclude that LZ\,Cep is in a transitory state where both components are close to entirely fill their Roche lobe. LZ\,Cep is close to be a contact binary.


\section{Stellar evolution and binarity}
\label{s_disc}

\begin{figure}[t]
\centering
\includegraphics[width=8cm]{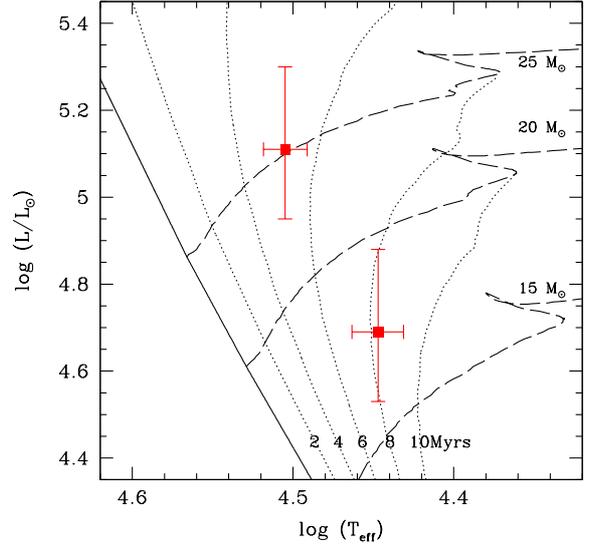}
\caption{HR diagram with the position of the two components of the LZ\,Cep system shown by the red squares. The primary is the most luminous star. The evolutionary tracks computed for an initial rotational velocity of 300 \kms~and the isochrones are from \citet{mm03}. }\label{fig_hr}
\end{figure}

The position of the two components of the LZ\,Cep system in the HR diagram is displayed in Fig.\ \ref{fig_hr}. The two stars are located significantly off the zero-age main sequence. The isochrones reveal that the primary would be assigned an age of 4--6 Myrs by using single star evolutionary tracks. The secondary would have an age of 7--9 Myrs. The distinct ages could be due to a different rotation rate. Indeed, as recently demonstrated by \citet{brott11}, isochrones based on rapidly rotating evolutionary models tend to be located closer to the ZAMS compared to isochrones based on tracks with lower rotation rates. The effect is present for stars with luminosities above 3.0 10$^{5}$ $L_{\odot}$ in the Galaxy. But it is significant only for values of \vsini\ larger than $\sim$250 \kms. Since the projected rotation rates we derive are 130 and 80 \kms for the primary and secondary components, we can exclude rotation as an explanation to the age difference between both stars.

We have determined various masses for the components of the LZ\,Cep system. From the orbital solution and the light curve analysis, we have derived the true present masses: about 16~\msun~and 6.5~\msun~for the primary and the secondary components, respectively. Those masses can be compared to the spectroscopic masses obtained from the luminosity, temperature and gravity. We obtained 15.9$^{+9.8}_{-8.4}$ \msun~for the primary and 4.1$^{+2.4}_{-2.1}$ \msun~for the secondary. Within the error bars, the agreement is very good for the primary, and marginal for the secondary. This could be due to the deformation of the secondary (see Fig.\ \ref{fit_lc}) and the fact that we derive only an average gravity. The evolutionary masses derived using single star evolutionary tracks are respectively 25.3$^{+6.2}_{-4.0}$ \msun~and 18.0$^{+2.4}_{-2.5}$~\msun~for the primary and the secondary. The former are in strong conflict with the latter. Under normal circumstances, a star of initial mass around 25~\msun\ does not loose 10~\msun\ through stellar winds before it enters the Wolf--Rayet phase \citep[e.g.][]{mm00}. So normal evolution cannot explain the mass discrepancy for the primary.  In addition, if the discrepancy concerned only one component, one could argue that mass was lost through Roche--lobe overflow and given to the other component. But in our case both stars have evolutionary masses much larger than present masses. This is an indirect indication that the system was affected by binary interaction and that single--star evolutionary tracks cannot be used blindly to derive the initial masses of the components of LZ\,Cep. We also note that the spectroscopic masses we obtained are lower than what is expected from single-star evolution \citep[see e.g.][]{mar05}. This is especially true for the secondary, for which a mass of 4.1$^{+2.4}_{-2.1}$~\msun\ is much lower than the 10-15~\msun\ of \citet{mar05}. Having such a low mass but still the luminosity and effective temperature of an O star imply the occurrence of processes not present in single-star evolution.

\begin{figure}[t]
\centering
\includegraphics[width=8cm]{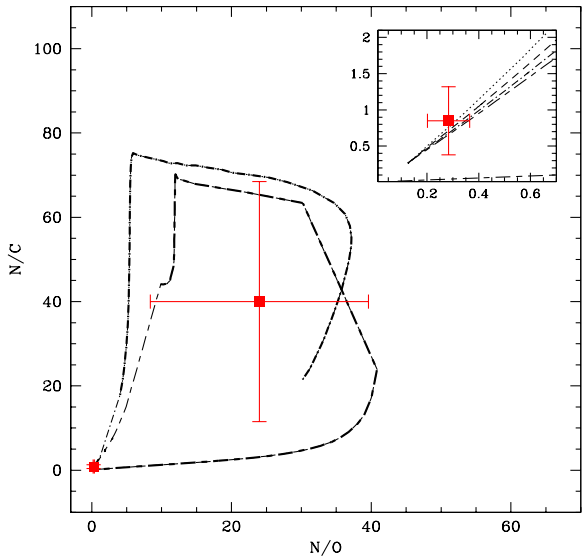}
\caption{N/C versus N/O for the components of LZ Cep (red squares). The inset zooms on the primary value. The evolutionary tracks are from \citet{mm03}. The dotted line (resp. dashed, dot--dashed, short--dash long--dashed lines) are tracks for a 15 \msun\ (resp. 20, 25, 40~\msun) star. The bold lines indicate the Wolf-Rayet phase of the evolution according to \citet{mm03}, i.e. X(H) $<$ 0.4 and \teff~$>$ 10000 K.}\label{fig_cno}
\end{figure}

Another indication of interaction within the system comes from the analysis of the surface abundances. In Fig.\ \ref{fig_cno} we show the nitrogen over carbon ratio as a function of the nitrogen over oxygen. Again, single-star evolutionary tracks including rotational mixing from \citet{mm03} are shown. We see that the primary, located at the bottom--left corner of Fig.\ \ref{fig_cno}, displays surface chemistry consistent with single star evolution. However, the secondary features extreme nitrogen enrichment and carbon/oxygen depletion. Such high N/C and N/O ratios are only reached in the Wolf-Rayet phase in single star evolution (the Wolf-Rayet phase is indicated by bold symbols in  Fig.\ \ref{fig_cno}). Interestingly, this is consistent with the high helium content of the secondary: $\rm{He/H}=0.4$, corresponding to $\rm{Y}=0.6$, is only observed in advanced states of evolution in single--star evolutionary tracks with initial masses of about 20~\msun. This raises the question of whether the secondary is not in fact more evolved than the primary and is currently in a state of central helium burning. Using the mass--luminosity relation of \citet{schaerer92} for core H--free stars, we find that a star with a current mass of about 6~\msun\ (resp. 5~\msun) should have a luminosity $\lL=4.85$ (resp. 4.69). This is remarkably consistent with the derived luminosity of the secondary ($\lL=4.69$). 

All indications thus favor the scenario in which the secondary was initially more massive, evolved faster, transferred mass to the companion and is now close to be a core He--burning object. This could also explain the rather low gravity and mass of the secondary (if it is truly a main-sequence O star, \logg\ should be much higher -- see also above). However, there are two caveats to this scenario. First, the spectroscopic appearance of the secondary is not that of an evolved massive star. With a spectral type ON\,9.7V, it looks like a normal main sequence star. Said differently, despite of all the indications, the absence of strong mass loss characterizing evolved massive stars (and affecting spectral types) is challenging. \citet{gh08} have shown that the mass loss rate of Wolf-Rayet stars depended on the Eddington factor $\Gamma_{e}$. Estimating this factor from the parameters we determined, we obtain $\Gamma_{e} \sim\ 0.17$. This is rather low, and much lower than the range of values for which \citet{gh08} provide mass loss prescriptions for Wolf--Rayet stars (their computations are restricted to $\Gamma_{e} > 0.4$). Hence, the secondary may very well have a rather normal wind. If we use the mass loss recipe of \citet{vink01}, we obtain $\log \dot{M} \sim -7.0$, which is much more typical of O stars. Consequently, it does not seem impossible that the secondary is an evolved object with a rather normal wind and thus the appearance of an O star. 

The second caveat is the chemical composition of the primary. If almost all the envelope of the secondary was dumped onto the primary, the chemical pattern should be different. The primary should show evidence for N enrichment and CO depletion since material from the secondary would be processed, the secondary being significantly evolved. According to our determinations, there are only weak signatures of CNO processing. This favors a weak accretion of material from the donor. A way out of this puzzle is to invoke a very non--conservative mass transfer even though this kind of process was not observed in the two post Roche lobe overflow key systems:  $\phi$\,per \citep{bozic1995} and RY\,Scuti \citep{grundstrom2007}. All in all, the scenario according to which the secondary was initially more massive and evolved faster to the core--He burning phase while filling its Roche lobe and transferring very inefficiently material onto the primary seems a valuable explanation of the system.

But other explanations can be invoked. \citet{demink07} and \citet{langer08} have argued that strong nitrogen enhancement can be obtained on interacting binaries. Two effects are at work: the mass gainer receives N--rich material usually from the inner layers of the donor; and the angular momentum received from the primary triggers rotational mixing which further increases the surface nitrogen content of the secondary. In total, at solar metallicities, the enrichment can be of a factor of ten compared to the initial N abundance. According to Table \ref{tab_orb}, the nitrogen surface abundance of the secondary star in LZ\,Cep is 25 times the initial value. This is qualitatively consistent with the predictions of binary evolution which depend on the orbital parameters and masses of the components. In that case, contrary to the scenario described above, the primary is the mass donor. Since rotational mixing is triggered by mass transfer, one also expects helium enrichment in the secondary, which is what we observe in the LZ\,Cep. However, according to \citet{dem09}, to reach a mass fraction $\rm{Y}=0.6$ requires very short periods and occurs only on stars more massive than $\sim$ 40~\msun. Another concern is that the rotational velocity of the secondary is not extreme. With \vsini\ = 80 \kms, and assuming that the rotation axis is parallel to the orbital axis, this gives a rotation rate of $\sim$ 100 \kms. But as shown by \citet{langer08}, the rotational velocity can reach very high values after mass transfer, and subsequently come back to more moderate values. The question is whether fast mixing can be triggered during this short spin--up phase. It seems unlikely, but further dedicated simulations should be run to tackle this problem. 

With the inclination, projected rotational velocities and radius, we can estimate the rotational period of each star and check if synchronization has been achieved in the system. Assuming an inclination angle of 49.1$^{\circ}$ (the average of the two values from Table \ref{tab_lc}), we obtain an orbital period of 3.09$\pm$0.24~d using R = 10.50 $R_{\odot}$ (average of the polar radii for the primary) and 3.75$\pm$0.31~d using R = 12.75 $R_{\odot}$ (equatorial radius). Similarly, the rotation period of the secondary is 3.15$\pm$0.45~d (polar radius) and 4.21$\pm$0.60~d (equatorial radius). Compared to the orbital period (3.07~d), the true rotational periods of both components (likely in between the polar and equatorial cases) are thus roughly 10 to 20\% larger, implying that the synchronous co-rotation is not completely established in the system. However, given the small differences and the similarity of the primary and secondary rotational periods, the system is certainly on the verge of achieving it.

All in all, we have thus gathered evidence for mass transfer and tidal interaction in the massive binary system LZ\,Cep. The scenario according to which the secondary was initially the more massive star appears to explain more observational constraints, although it is not free of challenges. What is clear is that single--star evolutionary tracks are certainly not relevant to explain the position of the two components in the HR diagram. This was highlighted by \citet{wellstein01} who computed evolutionary tracks for interacting binaries and showed that as soon as mass transfer occurs, dramatic deviations from single--star evolutionary track happen. Consequently, the evolutionary masses we derived in Table\ \ref{tab_orb} are very uncertain.

\section{Conclusion}
We have presented a detailed analysis of the binary system LZ\,Cep. We used high-resolution and high signal-to-noise spectra collected with the instrument NARVAL on the TBL. We performed an orbital solution from which we did not find evidence for a non-zero eccentricity. We have applied a spectral disentangling programme to our data set. This gave us the individual spectra of each component of the system. We have determined the spectral type of the system: O\,9III$+$ON\,9.7\,V. We have subsequently performed a quantitative spectroscopic analysis with the CMFGEN atmosphere code. Although the secondary star presents a main-sequence luminosity class, its surface abundances show a strong helium and nitrogen enhancement as well as carbon and oxygen depletion. The primary barely shows any sign of chemical processing. 
The analysis of the Hipparcos photometry has revealed ellipsoidal variations that can be explained by two configurations for this system: LZ\,Cep is a semi-detached system in which either the primary or the secondary almost entirely fills its Roche lobe. The ellipsoidal variations of the light curve correspond to an inclination of about 48$^{\circ}$. Together with the spectroscopic orbital solution, this allowed us to determine the masses of the system: about 16.0 \msun\ for the primary and 6.5 \msun\ for the secondary. These masses are in good agreement with the spectroscopic masses. The evolutionary masses obtained using single--star evolutionary tracks are larger. Our results indicate that LZ\,Cep is in a transitory state following mass transfer from the secondary -- which was initially the most massive component -- to the primary. The secondary is now (close to be) a core He--burning object. The mass transfer must have been very inefficient to account for the weak chemical processing at the surface of the primary. The caveats to this scenario have been discussed. The system is not yet completely synchronized, but the individual rotational periods are only 10 to 20\% larger than the orbital period.

Our investigation of the binary system LZ\,Cep represents a significant improvement compared to previous studies. We have obtained the individual spectra of each component. This leads to the determination of a number of stellar parameters which could be used to constrain the evolutionary status of the system. This evolved system is thus particularly interesting since it provides observational constraints to theoretical binary evolutionary models. Similar works exist for B stars \citep[e.g., see][]{pav09} but studies of systems composed of two evolved early-type stars are scarce. We have emphasized the need for theoretical binary system evolutionary models since comparison of the stellar properties of the components of evolved binary systems to single--star theoretical predictions revealed discrepancies.

\label{s_conc}

\begin{acknowledgements}
We thank the anonymous referee for valuable comments that helped to improve the presentation of the paper and the discussion of the results. LM thanks the PRODEX XMM/Integral contract (Belspo) and the Communaut\'e fran\c caise de Belgique -- Action de recherche concert\'ee -- A.R.C. -- Acad\'emie Wallonie-Europe for their support. We thank John Hillier for making his code CMFGEN available and for constant help with it.
\end{acknowledgements}

\bibliography{laurent.bib}



\end{document}